# Esperienze di didattica della fisica in diversi livelli del sistema educativo

**Abstract** La crescente partecipazione della società a eventi di divulgazione scientifica, i progetti sostenuti dal MIUR[1,2] per promuovere l'insegnamento delle materie STEM (Science, Technology, Engineering, Mathematics) nei diversi livelli del sistema educativo e l'introduzione di argomenti di fisica moderna e contemporanea nelle programmazioni di alcuni Licei[3] hanno contribuito, in molti casi, a rafforzare il legame tra scuola, Università e centri di ricerca. Questo legame si è concretizzato nell'istituzione di attività dedicate in cui è stato necessario impiegare metodologie comunicative e didattiche sempre più efficaci. L'articolo presenta lo spettro delle attività realizzate negli ultimi tre anni dall'Università di Ferrara e dall'Istituto Nazionale di Fisica Nucleare per la comunicazione e la didattica della fisica. Verranno analizzati alcuni casi studio che si differenziano per contenuti, destinatari, contesti e strategie didattiche. In particolare verranno prese in esame le attività dedicate all'insegnamento della fisica moderna e contemporanea condotte con gli allievi delle scuole secondarie di II grado [4], i laboratori scientifici hands-on realizzati con gli allievi delle scuole primarie e un'esperienza di didattica museale inserita in una mostra di storia della fisica dedicata al binomio arte-scienza[5]. Il primo è un caso di out-of-school learning in cui gli allievi delle scuole secondarie di II grado lavorano in gruppo a fianco dei ricercatori, per realizzare un esperimento di fisica moderna, e una volta tornati in classe devono relazionare ai pari quanto svolto e appreso durante l'attività laboratoriale. Nel secondo caso, gli allievi delle scuole primarie sono chiamati a condurre esperimenti guidati per acquisire familiarità con il metodo scientifico, investigando alcuni fenomeni fisici presenti nel quotidiano. Nell'ultimo caso, alcune scoperte della fisica moderna vengono introdotte dalla corrispondenza tra opere d'arte e strumenti scientifici e collegate alla storia di Ferrara.

The growing interest of people in science events, the projects supported by the Italian Ministry of Education, University and Research to foster STEM teaching in different levels of the education system and the introduction of modern physics in some Italian high schools, contributed to the strengthening of interaction between schools, universities and research centers. This interaction realized in dedicated activities characterized by innovative communication and education strategies. This paper presents the events of science dissemination organized in the last years by the University of Ferrara and the National Institute for Nuclear Physics taking into account some case study differentiated by contents, recipients and education strategies.

Autori: S. Bertelli [1][2], M. Andreotti [3], P. Lenisa [1], F. Spizzo [1]

   (1) Università degli Studi di Ferrara, Dipartimento di Fisica e Scienze della Terra
   (2) INFN, Laboratori Nazionali di Frascati
   (3) INFN, Sezione di Ferrara

Email: susanna.bertelli@unife.it

Susanna Bertelli, dottore di ricerca in Fisica e Master in Comunicazione delle Scienze è assegnista di ricerca in Fisica presso l'Università di Ferrara e i Laboratori Nazionali di Frascati dell'INFN e docente del corso di Didattica della Fisica. Si occupa di progetti di educazione scientifica e di orientamento per divulgare e insegnare la fisica nelle scuole e in programmi di educazione continua. E' ideatrice e responsabile del progetto Fisici Senza Frontiere, cura la Collezione Instrumentaria delle Scienze Fisiche del Sistema Museale d'Ateneo di Ferrara, fa parte della redazione del sito di divulgazione scientifica ScienzaPerTutti INFN, progetta e organizza mostre ed eventi scientifici. Ha vinto il bando Giovani Ricercatori dell'Università di Ferrara grazie al quale ha trascorso un periodo al CERN per studiarne le attività di outreach. E' autrice di pubblicazioni di fisica delle particelle e storia della Fisica.

Mirco Andreotti è Tecnologo di Ricerca presso la Sezione di Ferrara dell'Istituto Nazionale di Fisica Nucleare e collabora con le attività di ricerca e didattiche del Dipartimento di Fisica e Scienze della Terra dell'Università degli studi di Ferrara. Ha tenuto corsi di Fisica, Matematica, Informatica, Laboratori di Fisica ed Elettronica per diversi corsi di Laurea, attualmente è docente del Laboratorio di Elettronica Digitale per il corso di laurea in Fisica. La sua attività principale riguarda la progettazione, personalizzazione e applicazione della moderna tecnologica elettronica e informatica applicata ai laboratori di ricerca e ai laboratori didattici. In particolare collabora ad attività di ricerca dei gruppi di Fisica delle Particelle e di Fisica Medica. Ha inoltre collaborato ad attività di ricerca nei campi dell'architettura, della medicina e dell'ambiente per lo studio di tecniche di misura e analisi dati. E' autore di numerose pubblicazioni su riviste internazionali ed è stato relatore di numerose tesi di laurea in Informatica, Scienze Naturali e Fisica.

Paolo Lenisa ha conseguito il dottorato in Fisica presso l'Università di Ferrara dove è Professore Ordinario in Fisica Nucleare e Subnucleare e docente di Fisica Generale e Storia della Fisica. Si occupa di fisica delle particelle elementari con particolare riguardo allo studio dello spin del protone ed alle simmetrie fondamentali della Natura. In tale ambito, è stato proponente e responsabile di vari esperimenti in diversi laboratori internazionali. Attualmente è Co-spokesperson della collaborazione JEDI (Juelich Electric Dipole moment Investigations) che si si occupa del momento di dipolo elettrico del protone, una ricerca legata al mistero della dominanza della materia sull'antimateria nell'Universo. E' autore di oltre 200 pubblicazioni su rivista scientifica e relatore di 10 tesi di dottorato in Fisica.

Federico Spizzo è Ricercatore di Fisica Sperimentale presso il Dipartimento di Fisica e Scienze della Terra dell'Università degli Studi di Ferrara, ove tiene i corsi di 'Fisica dello Stato Solido' per il Corso di Laurea Magistrale in Fisica e di 'Fisica', per il Corso di Laurea in Scienze Biologiche. I sui principali interessi di ricerca riguardano lo studio dei materiali magnetici di tipo nanostrutturato. In quest'ambito, si occupa sia delle proprietà di sistemi magnetici biocompatibili che dello studio di materiali magnetici per memorizzazione dati e sensoristica. E' autore di numerose pubblicazioni su riviste internazionali, ed è stato relatore di oltre 30 tesi di laurea. Attualmente cura i 'Venerdì dell'Universo', rassegna di incontri dedicati alla divulgazione scientifica organizzata dal suo Dipartimento e dalla locale sezione dell'Istituto Nazionale di Fisica Nucleare.

## 1.La diffusione della cultura scientifica

Nell'ultimo decennio in Italia sono aumentate le attività per la promozione della cultura scientifica che si sono concretizzate in proposte e modelli innovativi sia nell'ambito dell'insegnamento che della divulgazione. Grazie a queste risorse, anche la didattica e la comunicazione della Fisica stanno vivendo uno sviluppo molto intenso. Basti pensare ai bandi attivati dal MIUR per la diffusione della cultura scientifica [Bando MIUR diffusione], al bando proposto dal Dipartimento per le pari opportunità per il finanziamento di percorsi di approfondimento in materie scientifiche con l'intento di superare stereotipi e pregiudizi che alimentano il gap di conoscenze tra le studentesse e gli studenti rispetto alle materie STEM [Bando Pari opportunità] e al Progetto Lauree Scientifiche [PLS]. Tra gli obiettivi principali di quest'ultimo progetto si ha quello di migliorare la conoscenza e la percezione delle discipline scientifiche nella Scuola secondaria di secondo grado, offrendo agli studenti degli ultimi tre anni di partecipare ad attività di laboratorio, curriculari ed extra curriculari stimolanti e coinvolgenti.
In ultimo, l'introduzione della Fisica moderna nelle programmazioni dei Licei Scientifici, in base alle Indicazioni Nazionali del MIUR, ha aperto un dialogo stimolante e costruttivo tra Scuole e Università che si interrogano in merito alle strategie, ai supporti e ai materiali didattici da adottare (Michelini, 2010).
In questo contributo verranno presentate le numerose iniziative messe a punto e realizzate dal Dipartimento di Fisica e Scienze della Terra dell'Università degli Studi di Ferrara in collaborazione con l'Istituto Nazionale di Fisica Nucleare (INFN), in merito alla disseminazione delle scienze fisiche con particolare attenzione alle ricadute didattiche. Tutte queste attività di trasferimento di conoscenze fanno parte della Terza Missione dell'Università e degli enti di ricerca [Anvur].

## 2. Out-of-school learning: laboratori di Fisica moderna all'Università
Il primo caso studio presentato è incluso nel programma delle attività di orientamento rivolte alle Scuole secondarie di II grado organizzate dal Corso di Laurea in Fisica a Ferrara.
Con Fisica moderna si indica quella parte del sapere scientifico nata in seguito alle teorie rivoluzionarie della meccanica quantistica e della relatività aprendo la strada allo sviluppo della fisica nucleare e subnucleare, della fisica della materia, l'astrofisica e la cosmologia.
Riguardo la programmazione di Fisica per l'ultimo anno dei Licei Scientifici, le Indicazioni Nazionali [Indicazioni Nazionali Licei Scientifici] riportano:

> Il percorso didattico comprenderà le conoscenze sviluppate nel XX secolo relative al microcosmo e al macrocosmo, accostando le problematiche che storicamente hanno portato ai nuovi concetti di spazio e tempo, massa ed energia. L'insegnante dovrà prestare attenzione a utilizzare un formalismo matematico accessibile agli studenti, ponendo sempre in evidenza i concetti fondanti.
>
> Lo studio della teoria della relatività ristretta di Einstein porterà lo studente a confrontarsi con la simultaneità degli eventi, la dilatazione dei tempi e la contrazione delle lunghezze; l'aver affrontato l'equivalenza massa-energia gli permetter. di sviluppare un'interpretazione energetica dei fenomeni nucleari (radioattività., fissione, fusione).
>
> L'affermarsi del modello del quanto di luce potrà essere introdotto attraverso lo studio della radiazione termica e dell'ipotesi di Planck (affrontati anche solo in modo qualitativo), e sarà sviluppato da un lato con lo studio dell'effetto fotoelettrico e della sua interpretazione da parte di Einstein, e dall'altro lato con la discussione delle teorie e dei risultati sperimentali che evidenziano la presenza di livelli energetici discreti nell'atomo. L'evidenza sperimentale della natura ondulatoria della materia, postulata da De Broglie, ed il principio di indeterminazione potrebbero concludere il percorso in modo significativo.
>
> La dimensione sperimentale potrà essere ulteriormente approfondita con attività da svolgersi non solo nel laboratorio didattico della scuola, ma anche presso laboratori di Università ed enti di ricerca, aderendo anche a progetti di orientamento.
>
> In quest'ambito, lo studente potrà approfondire tematiche di suo interesse, accostandosi alle scoperte più recenti della fisica (per esempio nel campo dell'astrofisica e della cosmologia, o nel campo della fisica delle particelle) o approfondendo i rapporti tra scienza e tecnologia (per esempio la tematica dell'energia nucleare, per acquisire i termini scientifici utili ad accostarsi criticamente il dibattito attuale, o dei semiconduttori, per comprendere le tecnologie più attuali anche in relazione a ricadute sul problema delle risorse energetiche, o delle micro- e nanotecnologie per lo sviluppo di nuovi materiali).

Come si evince dalle Indicazioni ci sono due aspetti fondamentali da tener presente nell'insegnamento della Fisica moderna: il formalismo matematico e la parte sperimentale. Gli allievi del quinto anno non possiedono tutti gli strumenti matematici per poter studiare e descrivere le leggi della fisica moderna e questo aspetto deve essere curato dal docente che deve adattare le spiegazioni al livello delle conoscenze degli allievi; l'aspetto matematico può essere integrato con un approccio storico-epistemologico. In questo approccio si mette in evidenza lo sviluppo del pensiero scientifico e dei concetti seguendo le tappe e le scoperte che fungono da pietra miliare per queste teorie. Per ciò che riguarda la parte sperimentale, non tutti i Licei hanno a disposizione attrezzature ed esperimenti dedicati alla parte di Fisica moderna.

In forza di quanto riportato dal MIUR, il Dipartimento di Fisica e Scienze della Terra dell'Università di Ferrara e l'INFN hanno istituito un'attività presso la propria sede "I laboratori di fisica moderna" dove poter approfondire l'aspetto sperimentale dei concetti di fisica moderna affrontati dagli allievi in classe.

Uno degli scopi di questa attività è avvicinare gli allievi al mondo della ricerca e in questa esperienza rivestono il ruolo di un ricercatore che lavora con i colleghi, esegue esperimenti, analizza e discute i risultati.

Vengono proposte dodici attività, ognuna delle quali è dedicata a uno dei temi presenti nelle Indicazioni. Alcune di queste attività sono state messe a punto ad hoc, altre sono state ottenute progettando delle trasposizioni didattiche delle attività di ricerca che sono realizzate in Dipartimento, adattando contenuti ed esperimenti per renderli fruibili a studenti del quinto anno. Gli allievi vengono suddivisi in gruppi e, una volta fornite loro le istruzioni, conducono in prima persona l'esperimento. L'approccio utilizzato è il *learning-by-doing* in quanto lo studente è parte attiva nell'esecuzione dell'esperienza e deve confrontarsi con gli altri componenti del gruppo sviluppando competenze nel *team working*. Spesso i gruppi di lavoro sono costituiti da studenti provenienti da diversi Istituti e quindi si cerca di privilegiare la strategia didattica del *cooperative learning* in cui si trovano a confrontare e condividere le loro conoscenze e le loro abilità con il comune obiettivo di completare l'esperimento. Gli allievi sono chiamati a osservare fenomeni e interpretarli, ad analizzare i dati raccolti, discutere i risultati ottenuti e dove possibile confrontarli con i modelli teorici. In questa attività vengono introdotti all'ambiente di ricerca dell'Università, imparano ad utilizzare attrezzature da laboratorio ad alto contenuto tecnologico, software e programmi di analisi dati e lavorano a fianco dei ricercatori. Una volta tornati in classe, viene adottata la strategia del *learning-by-teaching* in quanto gli studenti devono relazionare ai pari l'esperimento condotto in tutte le fasi: teoria, descrizione dell'apparato sperimentale, esecuzione e discussione dei risultati ottenuti. In questa attività gli allievi sono chiamati a ripensare e rielaborare quanto appreso e produrre una sintesi per i colleghi. Verranno di seguito introdotte le esperienze proposte agli allievi per presentare le diverse aree coinvolte e mettere in evidenza le conoscenze e le competenze che vengono sviluppate dai partecipanti [Laboratori Fisica Moderna].

Le prime quattro attività sperimentale sono dedicate allo studio della teoria di Planck, all'effetto fotoelettrico e al principio di Indeterminazione. Le restanti attività riguardano gli altri argomenti di fisica contemporanea citati nelle Indicazioni.

*Verifica del Principio di Indeterminazione di Heisenberg*
L'esperienza è finalizzata alla verifica del principio di indeterminazione di Heisenberg, uno dei concetti cardine della meccanica quantistica nominato nelle Indicazioni.
Nel laboratorio di ottica, tramite l'utilizzo di un laser e di una fenditura regolabile, gli allievi sono chiamati a interpretare le figure di diffrazione in termini del dualismo onda-corpuscolo della radiazione elettromagnetica. Per la verifica del principio di indeterminazione l'esperienza si propone di misurare le incertezze sulla posizione e sulla velocità trasversali del fascio laser. Gli studenti saranno guidati nelle operazioni di calibrazione degli strumenti, nell'esecuzione delle misure e nell'elaborazione dati.

*Misura della costante di Planck usando Diodi Emettitori di Luce (LED)*
La costante di Planck costituisce un parametro fondamentale che interviene nello studio dei fenomeni quantistici. Dopo una introduzione alla fisica moderna e alla tecnologia LED, gli allievi misurano e rappresentano graficamente la caratteristica tensione-corrente di diversi dispositivi commerciali, arrivando a stimare la tensione di soglia di ciascuno di essi. Lo studio della dipendenza della tensione di soglia dalla lunghezza d'onda permetterà la determinazione del valore della costante di Planck. Durante l'analisi si confronteranno i dati raccolti con i modelli, prestando attenzione alla trattazione consistente degli errori sperimentali.

*Effetto fotoelettrico e misura della costante di Planck*
Anche in questo caso, dopo un'introduzione alla fisica moderna, viene presentato l'effetto fotoelettrico e i problemi che lo studio di questo fenomeno mette in evidenza in ambito classico. Gli studenti procederanno con la misura della corrente emessa da un metallo quando, sotto vuoto, viene colpito da radiazione elettromagnetica e in particolare misureranno l'energia degli elettroni emessi. Tale misura verrà fatta utilizzando sorgenti di diversa frequenza, e lo studio della dipendenza dell'energia degli elettroni emessi dalla frequenza consentirà loro di stimare il valore della costante di Planck.

*L'effetto fotovoltaico nei semiconduttori*
Questo laboratorio è dedicato allo studio della fisica dei materiali. Gli studenti svolgono un'esperienza finalizzata all'osservazione dell'effetto fotovoltaico nei materiali semiconduttori. Verranno presentati i livelli energetici degli elettroni all'interno dei solidi cristallini, soffermandosi sul caso particolare dei semiconduttori; verrà poi descritta l'interazione tra la luce ed il materiale di tipo semiconduttore, interazione che è alla base del principio di funzionamento delle celle fotovoltaiche. In seguito, si svolgeranno misure di corrente generata da celle fotovoltaiche costruite usando diversi materiali di tipo semiconduttore, per osservare come la differente struttura microscopica dei livelli elettronici di ciascun semiconduttore influenzi la risposta macroscopica della cella.

*Focalizzazione dei raggi X, una lente per lo spazio*
Nell'esperienza di laboratorio gli studenti studiano un apparato per lo studio dei raggi X provenienti dalle stelle, partecipando alle misure di diffrazione dei raggi X attraverso delle tessere di materiale cristallino di elementi come Silicio, Germanio, Gallio. Questi elementi quando vengono investiti da radiazione producono il fenomeno della diffrazione e quando opportunamente orientati danno luogo ad un effetto cumulativo che risulta in una grande quantità di raggi X concentrati nel cosiddetto fuoco della lente. Gli studenti valutano il potere penetrante dei raggi X attraverso radiografie di oggetti comuni, misurano l'energia dei raggi X diffratti dai tasselli cristallini e ne vedono l'immagine con i rivelatori in dotazione al laboratorio. Vengono poi calcolate alcune proprietà di ciascuno dei cristalli in uso da confrontare con le proprietà teoriche richieste dal progetto di costruzione della lente spaziale. Le tessere che rispondo a certi criteri vengono selezionate per il prototipo da testare.

*Applicazioni della fluorescenza allo studio di opere d'arte*
In questo percorso gli allievi studiano le tecnologie nucleari applicate alla conservazione dei beni culturali, mettendo in evidenza la natura trasversale della Fisica.
Il fenomeno della fluorescenza, ben compreso e interpretato nel quadro della Fisica Quantistica, è alla base di diverse tecniche per lo studio di materiali. La caratteristica comune di queste tecniche è il fatto di non essere invasive, aspetto che le rende particolarmente apprezzate quando gli oggetti di studio sono opere d'arte. Verranno presentati esempi di fluorescenza da luce ultravioletta, suggestivi perché osservabili direttamente a occhio nudo, e verranno osservati gli spettri di raggi X prodotti per fluorescenza da pigmenti e da altri materiali artistici. Gli studenti imparano a riconoscere un pigmento antico in base allo spettro X rivelato.

*Nanostrutture di semiconduttori*
Partendo dalle conoscenze di base di fisica moderna vengono introdotti i concetti fondamentali della teoria dei livelli elettronici nei solidi cristallini. Vengono descritte le proprietà delle nano strutture (grani) di semiconduttori ed eseguite esperienze di misure elettriche sui sensori con vari gas ed una misura della barriera di potenziale tra grano e grano. Per comprendere le applicazioni al quotidiano viene illustrato il funzionamento di alcuni dispositivi realizzati presso il laboratorio di sensori del Dipartimento per le misure di inquinanti atmosferici in campo e per l'analisi del respiro di soggetti sani e malati.

*Radioattività naturale*
Nell'opinione pubblica la parola radioattività assume spesso un'accezione negativa: ci si dimentica che essa è un fenomeno fisico naturale, le cui applicazioni tecnologiche hanno portato a straordinari miglioramenti della qualità della vita dell'umanità. In questa attività la radioattività viene raccontata partendo dagli esperimenti di fisica fondamentale più complessi realizzati nei laboratori sotterranei.

*Magnetismo e nanostrutture*
Lo scopo dell'esperienza è quello di presentare agli studenti quali siano le peculiarità delle nanostrutture magnetiche e come sia possibile caratterizzarne le proprietà magnetiche. Dopo una breve introduzione, in cui si descrive quale sia la struttura dei livelli elettronici dei solidi e come da esse dipendano le proprietà dei materiali ferromagnetici, verrà mostrato come la dimensione nanometrica dei materiali possa influire sulle loro proprietà. Verrà poi presentata agli studenti la metodologia di caratterizzazione di tali materiali, metodologia basata sull'interazione radiazione-materia. In laboratorio viene effettuata la misura del ciclo di isteresi per verificare quanto presentato nella parte introduttiva. In alcuni casi verranno caratterizzati materiali già disponibili in laboratorio, in altri casi si procederà prima alla produzione e poi alla caratterizzazione dei materiali nanostrutturati.

*Laboratorio di eco-fluidodinamica e applicazioni in medicina*
L'attività sperimentale inizia con una parte introduttiva sulla fisica degli ultrasuoni e sul loro utilizzo in campo diagnostico e proseguirà con alcune prove pratiche svolte dagli studenti tra le quali: misura dell'attenuazione di un'onda ultrasonora al variare delle caratteristiche del mezzo in cui si propaga l'onda e della frequenza nominale dell'onda stessa; studio del flusso di un liquido tramite ecoDoppler: utilizzando un apparato che permette di riprodurre la circolazione sanguigna in carotide e giugulare verranno effettuate misure di flusso tramite acquisizione di profili Doppler ed immagini Brightness-Mode (Sisini, 2015).

*Laboratorio di Cosmologia*
Il percorso mira a fornire agli studenti delle nozioni base di Cosmologia, con enfasi sulla radiazione cosmica di fondo, che è la radiazione elettromagnetica prodotta nell'Universo primordiale il cui residuo "fossile" permea l'universo. Scoperta nel 1964, è la maggiore evidenza sperimentale del modello del Big Bang.

Alle lezioni teoriche seguono lezioni pratiche dove gli studenti analizzano i dati del satellite Planck dell'ESA con lo scopo di vincolare alcune proprietà del modello cosmologico standard. Come proposto nelle indicazioni, lo studente avrà l'opportunità di accostarsi alle scoperte più recenti nell'ambito della Cosmologia confrontandosi direttamente con i ricercatori specializzati in questo campo.

*Laboratorio di Fisica Teorica*
Le leggi della fisica, dalla meccanica classica all'elettromagnetismo fino alla meccanica quantistica sono espresse come "equazioni differenziali". Dopo una breve e semplice introduzione teorica a questo tipo di strumento matematico, usando un software dedicato verranno risolte al computer alcune equazioni differenziali già note agli studenti (come la legge di gravitazione universale di Newton) ed alcune semplici equazioni della fisica moderna (esempi tratti dalla relatività di Einstein). Gli studenti lavorano direttamente col software e sono guidati nella risoluzione dei problemi e nella visualizzazione e analisi dei risultati.

Gli allievi in questa attività hanno l'opportunità di apprendere alcuni concetti di fisica moderna fuori dall'ambiente scolastico, all'interno di un ambiente di ricerca, con attività *hands-on* non mediate da altre persone ma svolte direttamente da loro. Agli studenti viene richiesto di costruire il sapere attraverso l'esperimento avendo la possibilità di confrontarsi con i ricercatori. Oltre l'aspetto didattico prettamente legato al programma di Fisica del quinto anno, gli studenti possono vivere un'esperienza di orientamento universitario e quindi interrogarsi sulle proprie vocazioni. Dal punto di vista dei ricercatori e dell'Università questa è un'esperienza unica per disseminare la loro ricerca entrando in contatto con i docenti, gli allievi e loro famiglie.

### 3. Out-of-school learning: eventi di divulgazione scientifica

In questa sezione vengono descritte le diverse iniziative inserite nel contesto dell'apprendimento informale con particolare riguardo a quelle che avvengono al di fuori dell'ambiente scolastico in Università e centri di ricerca, che fungono da catalizzatore per l'insegnamento/apprendimento della Fisica. Questi eventi costituiscono un ponte tra il mondo della ricerca e la società portando in molti casi alla realizzazione di progetti di educazione scientifica che coinvolgono docenti e allievi delle scuole primarie e secondarie.
La prima iniziativa presa in esame è "Porte Aperte a Fisica" nata nel 2000 nella sede dell'ex Dipartimento di Fisica di Via Paradiso. L'evento, strutturato in sette giorni, aveva lo scopo di creare uno spazio dove far conoscere al grande pubblico le attività di ricerca portate avanti nell'ambito della Fisica a Ferrara. Questa manifestazione costituisce uno degli esempi di *best practice* della disseminazione della scienza sviluppati nell'Ateneo ferrarese in quanto il suo pattern dal 2011 è stato esteso alle attività dell'area scientifico-tecnologica divenendo "Porte Aperte al Polo Scientifico Tecnologico". L'evento è oggi caratterizzato da due percorsi di visita, senior e junior, differenziati in base all'età dei partecipanti. Nel percorso senior le persone assistono alla presentazione delle attività di ricerca nei laboratori da parte dei ricercatori. Particolare attenzione è rivolta agli allievi e ai docenti delle scuole che hanno la possibilità di vedere ambienti e laboratori e conoscere i protagonisti della ricerca. Dal punto di vista didattico questa esperienza costituisce un'opportunità per gli studenti di vivere l'attualità scientifica e per i docenti di rimanere aggiornati sui principali temi della ricerca, proponendo dove possibile approfondimenti in classe. Per quello che riguarda l'area Fisica vengono illustrate le attività di fisica della materia, sensori, applicazioni mediche, energie rinnovabili, fisica applicata alla medicina, astrofisica, tecnologie nucleari applicate all'ambiente e alla conservazione dei beni culturali, le ricerche di frontiera attraverso le attrezzature sviluppate per investigare la fisica delle particelle elementari e la fisica dello spazio. Tutti questi argomenti sono inseriti nelle Indicazioni Nazionali del MIUR.
Nel percorso junior, i bambini dai 5 a 11 anni assistono a interventi ludico-didattici dedicati alla scienza. Dal percorso junior fisica è nato come *by-product* nel 2014 un progetto di educazione scientifica per insegnare la fisica nelle scuole primarie e secondarie di I grado "Fisici Senza Frontiere", che verrà descritto nelle prossime sezioni.

All'interno di Porte Aperte 2016 è stato realizzato un progetto di Alternanza Scuola Lavoro [Alternanza Scuola Lavoro] che ha coinvolto alcuni studenti del Liceo Scientifico A. Roiti di Ferrara in un percorso di comunicazione della scienza, fornendo supporto al personale e ai ricercatori coinvolti nell'organizzazione dell'evento.

Altre attività di divulgazione che sono da supporto e incentivo per la didattica sono La Notte dei Ricercatori e i Venerdì dell'Universo [Venerdì dell'Universo]. La Notte dei Ricercatori propone la presentazione delle attività di ricerca fuori dagli ambienti universitari per stabilire una forte connessione con il territorio abbattendo il pregiudizio che gli scienziati vedano i laboratori come torri d'avorio onorando un patto sociale per arricchire la società e rendere consapevoli i cittadini dei progressi della scienza e di come utilizzarli nelle scelte di tutti i giorni. I Venerdì dell'Universo, nati nel 2000, propongono seminari divulgativi dedicati alla scienza, alla sua applicazione in diverse discipline mettendo in rilievo i benefici per i cittadini.

Viene ora preso in esame un caso studio di didattica museale legato alla mostra di strumenti storici *"Fisica e Metafisica? La Scienza ai tempi di de Chirico e Carrà"* che si è tenuta a Palazzo Turchi di Bagno a Ferrara, dal novembre 2015 a gennaio 2016, organizzata dal Dipartimento di Fisica e Scienze della Terra, dall'INFN e dal Sistema Museale d'Ateneo dell'Università di Ferrara. L'evento è stato realizzato allo scopo di valorizzare il patrimonio storico scientifico dell'Ateneo coniugando arte, scienza e storia di Ferrara.

La mostra scientifica ha ripercorso le tappe principali dello sviluppo della Fisica negli anni tra fine Ottocento e inizio Novecento mettendo in luce gli sviluppi e le ricerche nel territorio ferrarese, attraverso l'esposizione di strumenti appartenuti in quegli anni al Gabinetto di Fisica e all'Osservatorio meteorologico (Zini, 2004, Caracciolo, 2009), quando la loro direzione era affidata a Giuseppe Bongiovanni e oggi conservati nella Collezione Instrumentaria delle Scienze Fisiche, sezione del Sistema Museale d'Ateneo.

Giuseppe Bongiovanni, professore di Fisica sperimentale, frequentava il gruppo di intellettuali composto da Giorgio de Chirico, Alberto Savinio, Carlo Carrà e Filippo de Pisis che si creò a Ferrara durante gli anni della Prima Guerra Mondiale. Il legame di amicizia si evince dalla presenza di Bongiovanni, nominato affettuosamente l'astronomo, nella prosa e nelle poesie di de Chirico, Savinio e de Pisis. Questo legame ha ispirato la mostra di strumenti storici che è stata inserita tra gli eventi collaterali della mostra d'arte dedicata a de Chirico e alle avanguardie metafisiche di Palazzo Diamanti, creando un legame molto forte tra Università e territorio.

Bongiovanni fu personaggio molto noto nell'ambiente scientifico italiano e internazionale dell'epoca. Si occupò di ricerca in diversi campi della Fisica e si dedicò con passione all'insegnamento. Membro di diverse Accademie scientifiche nazionali e internazionali, nel 1897 fu tra i firmatari della circolare che portò alla costituzione della Società Italiana di Fisica (Graziani Bottoni, 2000). Realizzò uno studio dettagliato sul clima di Ferrara basato su molti anni di osservazioni che pubblicò nel 1900 (Bongiovanni, 1900).

Il filo conduttore della mostra è il *divertissment* di stabilire una connessione tra la fisica e l'arte metafisica attraverso la lettura soggettiva che uno scienziato fornisce di queste opere. L'esposizione è stata strutturata in cinque sezioni: misure e campioni di misure, meteorologia, elettromagnetismo, astronomia e apparati medicali. Ogni sezione è abbinata a opere metafisiche e documenti dell'epoca. Dall'osservazione di alcuni particolari presenti nei quadri, con un gioco di corrispondenze, vengono introdotti gli strumenti scientifici e illustrate le principali scoperte che questi hanno determinato. Alle scuole in visita è stato proposto un approfondimento tecnologico riguardo le attrezzature scientifiche per capire come state realizzate le scoperte che hanno gettato le basi alla fisica moderna, presentando anche i personaggi e gli scienziati che hanno animato queste scoperte curando al contempo l'aspetto umano della scienza nel racconto delle vicende della vita di queste persone.

Nell'aree relative a meteorologia, elettromagnetismo e apparati medicali sono stati esposti gli strumenti utilizzati da Giuseppe Bongiovanni e presentati in alcuni suoi lavori a stampa (Bongiovanni, 1898; Bongiovanni 1900). In particolare, sono stati esposti lo psicrometro a ventilatore per la misura dell'umidità dell'aria e alcuni termometri descritti da Bongiovanni nel rapporto sul clima. E' stato esposto il radiotelegrafo marconiano che veniva usato per inviare giornalmente i dati meteorologici da lui registrati e che rappresenta una delle prime applicazioni delle onde elettromagnetiche (scoperte da Hertz nel 1888) per la comunicazione. Uno spazio della mostra è stato dedicato al principio di funzionamento dei tubi a scarica e dei tubi di Crookes con i quali sono state studiate le proprietà degli elettroni e dei raggi X. Grazie alla presenza di questi strumenti è stata introdotta la scoperta dei raggi X ad opera di Willhelm Conrad Roentgen (1895) che ha cambiato radicalmente la nostra visione della natura facendoci accedere al mondo invisibile. Questa scoperta costituisce un esempio molto importante della fisica applicata al vivere quotidiano, perché già nel 1896 vennero impiegati per la radiodiagnostica in medicina e durante la Prima Guerra Mondiale gli ospedali da campo erano attrezzati con un reparto di radiologia.
Sono state organizzate visite guidate per le scuole e per la cittadinanza e per i più piccoli dei laboratori didattici interattivi di fisica.
L'evento ha consentito di presentare alle persone il patrimonio scientifico storico della Collezione di Fisica, avvicinando le persone alla scienza tramite l'arte e raccontando gli sviluppi tramite gli strumenti e le persone presentando le ricadute nel quotidiano e come queste scoperte furono vissute e studiate nella realtà ferrarese.

4. PROGETTO DI EDUCAZIONE SCIENTIFICA NELLE SCUOLE DI FERRARA

Come anticipato nei precedenti paragrafi, il percorso Porte Aperte Junior Fisica, all'interno della manifestazione di apertura al pubblico del Polo Scientifico Tecnologico, ha instaurato un contatto diretto tra Università e Scuole primarie e secondarie di I grado di Ferrara, evidenziando come i docenti spesso non si sentano sicuri nell'insegnare le materie scientifiche perché non proprie della loro formazione. Da un dialogo aperto con i docenti della scuola primaria è nato nel 2014 un progetto di educazione scientifica "Fisici Senza Frontiere" con lo scopo di proporre degli interventi didattici laboratoriali nelle scuole con temi inerenti alla fisica generale. Ogni lezione è dedicata ad un argomento come ottica, calore e temperatura, elettricità, meteorologia, pressione e vuoto, astronomia. L'intervento didattico alterna discussioni guidate ad attività pratiche svolte dagli allievi individualmente o in gruppo. La classe viene solitamente divisa in gruppi e viene chiesto agli allievi di collaborare e di confrontare le idee. Si cerca di privilegiare l'apprendimento per immagini e l'apprendimento cinestetico. Seguendo il metodo scientifico, gli allievi realizzano esperimenti e discutono i risultati ottenuti.
Vengono studiati esempi e controesempi tenendo uno sguardo rivolto all'esperienza quotidiana per far capire agli allievi che quanto appreso è per loro uno strumento per interpretare i fenomeni che li circondano. Le esperienze dimostrative realizzate dai tutor sono condotte utilizzando sia materiale low tech facilmente reperibile che attrezzature scientifiche da laboratorio, portate appositamente nelle scuole per poter far osservare alcuni fenomeni in prima persona dagli allievi. Ne è un esempio il percorso "pressione e vuoto" in cui si porta un sistema da vuoto per creare in maniera efficace il vuoto dentro un recipiente e osservare cosa succede ad alcuni fenomeni quando viene tolta l'aria (caduta di oggetti nel tubo di Newton, palloncino gonfio, acqua che bolle a temperatura ambiente, marshmallow che viene deformato) o il percorso di ottica in cui si utilizza un kit didattico di ottica geometrica per studiare e verificare le sue leggi.
Come esempio viene presentato in dettaglio il percorso "Calore e temperatura", incentrato sui concetti chiave della termologia per introdurre agli allievi la differenza tra calore e temperatura, conduttori e isolanti termici e i metodi di propagazione del calore. In questa unità gli allievi imparano attraverso un'esperienza sensoriale la differenza tra conduttore e isolante termico. Dopo aver spiegato la conduzione termica, vengono introdotti materiali ed esempi e con l'inquiry based learning attraverso domande stimolo, gli allievi devono interpretare alcuni fenomeni utilizzando schemi e disegni. Viene introdotta la grandezza temperatura, viene spiegato il principio di funzionamento del termometro e la sua lettura. Come attività sperimentale gli allievi sono chiamati a costruire un termoscopio e spiegare come funzionano alcuni giochi scientifici. Vengono presi in esame i fenomeni riguardanti i passaggi di stato e la dilatazione termica nei solidi, liquidi e gas. Infine vengono introdotti i metodi di propagazione del calore attraverso l'osservazione di alcune esperienze realizzate con materiale facilmente reperibile. Gli allievi sono parte attiva nella costruzione del percorso e sono previsti dei momenti di ricapitolazione in cui loro stessi raccontano cosa hanno appreso.
Al termine della lezione viene proposto un gioco sotto forma di mappa concettuale o cruciverba per ordinare e creare una sintesi della lezione.
Durante l'anno scolastico 2016/2017 sono stati fatti interventi didattici in sedici scuole della provincia di Ferrara, Milano e Frascati. Grazie a questa attività si è creato un forte legame tra scuole e Università per promuovere la *scientific literacy* dal primo ciclo d'istruzione.

[Indicazioni Nazionali Licei Scientifici] Indicazioni Nazionali Licei Scientifici URL: http://www.indire.it/lucabas/lkmw_file/licei2010///indicazioni_nuovo_impaginato/_Liceo%20scientifico%20opzione%20Scienze%20Applicate.pdf

[Laboratori Fisica Moderna] Laboratori di Fisica Moderna organizzati dal Corso di Laurea in Fisica, Università di Ferrara URL: http://www.fe.infn.it/orientamento_fisica/courses/laboratori-di-fisica-moderna/

[Alternanza Scuola Lavoro] Alternanza Scuola Lavoro URL:http://www.istruzione.it/alternanza/index.shtml

[Venerdì dell'Universo] I Venerdì dell'Universo URL: http://www.fe.infn.it/venerdi/